\documentstyle[10pt,epsf,epsfig,hangcaption,xspace,amssymb,amsfonts,amsmath,amsthm,cite,
dp_delphititle,lineno]{dp_delphi}
\makeindex
\def\DpPaperGroup{EP} 
\def\DpPaperRef{2003-087}
\def\DpDate{10 December 2003}
\def\DpAuthors{DELPHI Collaboration}
\def\DpSubmit{(Accepted by Euro. Phys. J. C)}
\def\DpTitle{{ 
Search for fermiophobic Higgs bosons in final states with photons at LEP 2}}
\def\DpComment{  }
\def\DpEMail{  }
\newcommand{\gev}{{\ifmmode \mbox{Ge\kern-0.2exV}
\else Ge\kern-0.2exV\nolinebreak\fi}}
\newcommand{\mev}{{\ifmmode \mbox{Me\kern-0.2exV}
\else Me\kern-0.2exV\nolinebreak\fi}}

\begin{document}
\makeatletter
\newcount\@tempcntc
\def\@citex[#1]#2{\if@filesw\immediate\write\@auxout{\string\citation{#2}}\fi
  \@tempcnta\z@\@tempcntb\m@ne\def\@citea{}\@cite{\@for\@citeb:=#2\do
    {\@ifundefined
       {b@\@citeb}{\@citeo\@tempcntb\m@ne\@citea\def\@citea{,}{\bf ?}\@warning
       {Citation `\@citeb' on page \thepage \space undefined}}%
    {\setbox\z@\hbox{\global\@tempcntc0\csname b@\@citeb\endcsname\relax}%
     \ifnum\@tempcntc=\z@ \@citeo\@tempcntb\m@ne
       \@citea\def\@citea{,}\hbox{\csname b@\@citeb\endcsname}%
     \else
      \advance\@tempcntb\@ne
      \ifnum\@tempcntb=\@tempcntc
      \else\advance\@tempcntb\m@ne\@citeo
      \@tempcnta\@tempcntc\@tempcntb\@tempcntc\fi\fi}}\@citeo}{#1}}
\def\@citeo{\ifnum\@tempcnta>\@tempcntb\else\@citea\def\@citea{,}%
  \ifnum\@tempcnta=\@tempcntb\the\@tempcnta\else
   {\advance\@tempcnta\@ne\ifnum\@tempcnta=\@tempcntb \else \def\@citea{--}\fi
    \advance\@tempcnta\m@ne\the\@tempcnta\@citea\the\@tempcntb}\fi\fi}
 
\makeatother
\begin{titlepage}
\pagenumbering{roman}
\CERNpreprint{\DpPaperGroup}{\DpPaperRef} 
\date{{\small\DpDate}} 
\title{\DpTitle} 
\address{\DpAuthors} 
\begin{shortabs} 
\noindent
Higgs boson production with subsequent decay to photons was searched for in
the data collected  by the DELPHI detector at centre-of-mass energies between 
183 GeV and 209 GeV, 
corresponding to a total integrated luminosity of nearly 650 pb$^{-1}$.
No evidence for a signal was found, and limits were set on 
$h^0Z^0$ and $h^0A^0$ production with $h^0$ decay to photons.
These results were used to exclude regions in the parameter space of 
fermiophobic scenarios of Two Higgs Doublet Models.
 
\end{shortabs}
\vfill
\begin{center}
\DpSubmit \ \\ 
\DpComment \ \\
\DpEMail \ \\
\end{center}
\vfill
\clearpage
\headsep 10.0pt
\addtolength{\textheight}{10mm}
\addtolength{\footskip}{-5mm}
\begingroup
%
\newcommand{\DpName}[2]{\hbox{#1$^{\ref{#2}}$},\hfill}
\newcommand{\DpNameTwo}[3]{\hbox{#1$^{\ref{#2},\ref{#3}}$},\hfill}
\newcommand{\DpNameThree}[4]{\hbox{#1$^{\ref{#2},\ref{#3},\ref{#4}}$},\hfill}
\newskip\Bigfill \Bigfill = 0pt plus 1000fill
\newcommand{\DpNameLast}[2]{\hbox{#1$^{\ref{#2}}$}\hspace{\Bigfill}}
%
\footnotesize
\noindent
\DpName{J.Abdallah}{LPNHE}
\DpName{P.Abreu}{LIP}
\DpName{W.Adam}{VIENNA}
\DpName{P.Adzic}{DEMOKRITOS}
\DpName{T.Albrecht}{KARLSRUHE}
\DpName{T.Alderweireld}{AIM}
\DpName{R.Alemany-Fernandez}{CERN}
\DpName{T.Allmendinger}{KARLSRUHE}
\DpName{P.P.Allport}{LIVERPOOL}
\DpName{U.Amaldi}{MILANO2}
% ...
\normalsize
\endgroup
\newpage
\titlefoot{Department of Physics and Astronomy, Iowa State
     University, Ames IA 50011-3160, USA
    \label{AMES}}
\titlefoot{Physics Department, Universiteit Antwerpen,
     Universiteitsplein 1, B-2610 Antwerpen, Belgium \\
     \indent~~and IIHE, ULB-VUB,
     Pleinlaan 2, B-1050 Brussels, Belgium \\
     \indent~~and Facult\'e des Sciences,
     Univ. de l'Etat Mons, Av. Maistriau 19, B-7000 Mons, Belgium
    \label{AIM}}
\titlefoot{Physics Laboratory, University of Athens, Solonos Str.
     104, GR-10680 Athens, Greece
    \label{ATHENS}}
\titlefoot{Department of Physics, University of Bergen,
     All\'egaten 55, NO-5007 Bergen, Norway
    \label{BERGEN}}
\titlefoot{Dipartimento di Fisica, Universit\`a di Bologna and INFN,
     Via Irnerio 46, IT-40126 Bologna, Italy
    \label{BOLOGNA}}
\titlefoot{Centro Brasileiro de Pesquisas F\'{\i}sicas, rua Xavier Sigaud 150,
     BR-22290 Rio de Janeiro, Brazil \\
     \indent~~and Depto. de F\'{\i}sica, Pont. Univ. Cat\'olica,
     C.P. 38071 BR-22453 Rio de Janeiro, Brazil \\
     \indent~~and Inst. de F\'{\i}sica, Univ. Estadual do Rio de Janeiro,
     rua S\~{a}o Francisco Xavier 524, Rio de Janeiro, Brazil
    \label{BRASIL}}
\titlefoot{Coll\`ege de France, Lab. de Physique Corpusculaire, IN2P3-CNRS,
     FR-75231 Paris Cedex 05, France
    \label{CDF}}
\titlefoot{CERN, CH-1211 Geneva 23, Switzerland
    \label{CERN}}
\titlefoot{Institut de Recherches Subatomiques, IN2P3 - CNRS/ULP - BP20,
     FR-67037 Strasbourg Cedex, France
    \label{CRN}}
\titlefoot{Now at DESY-Zeuthen, Platanenallee 6, D-15735 Zeuthen, Germany
    \label{DESY}}
\titlefoot{Institute of Nuclear Physics, N.C.S.R. Demokritos,
     P.O. Box 60228, GR-15310 Athens, Greece
    \label{DEMOKRITOS}}
\titlefoot{FZU, Inst. of Phys. of the C.A.S. High Energy Physics Division,
     Na Slovance 2, CZ-180 40, Praha 8, Czech Republic
    \label{FZU}}
\titlefoot{Dipartimento di Fisica, Universit\`a di Genova and INFN,
     Via Dodecaneso 33, IT-16146 Genova, Italy
    \label{GENOVA}}
\titlefoot{Institut des Sciences Nucl\'eaires, IN2P3-CNRS, Universit\'e
     de Grenoble 1, FR-38026 Grenoble Cedex, France
    \label{GRENOBLE}}
\titlefoot{Helsinki Institute of Physics, P.O. Box 64,
     FIN-00014 University of Helsinki, Finland
    \label{HELSINKI}}
\titlefoot{Joint Institute for Nuclear Research, Dubna, Head Post
     Office, P.O. Box 79, RU-101 000 Moscow, Russian Federation
    \label{JINR}}
\titlefoot{Institut f\"ur Experimentelle Kernphysik,
     Universit\"at Karlsruhe, Postfach 6980, DE-76128 Karlsruhe,
     Germany
    \label{KARLSRUHE}}
\titlefoot{Institute of Nuclear Physics PAN,Ul. Radzikowskiego 152,
     PL-31142 Krakow, Poland
    \label{KRAKOW1}}
\titlefoot{Faculty of Physics and Nuclear Techniques, University of Mining
     and Metallurgy, PL-30055 Krakow, Poland
    \label{KRAKOW2}}
\titlefoot{Universit\'e de Paris-Sud, Lab. de l'Acc\'el\'erateur
     Lin\'eaire, IN2P3-CNRS, B\^{a}t. 200, FR-91405 Orsay Cedex, France
    \label{LAL}}
\titlefoot{School of Physics and Chemistry, University of Lancaster,
     Lancaster LA1 4YB, UK
    \label{LANCASTER}}
\titlefoot{LIP, IST, FCUL - Av. Elias Garcia, 14-$1^{o}$,
     PT-1000 Lisboa Codex, Portugal
    \label{LIP}}
\titlefoot{Department of Physics, University of Liverpool, P.O.
     Box 147, Liverpool L69 3BX, UK
    \label{LIVERPOOL}}
\titlefoot{Dept. of Physics and Astronomy, Kelvin Building,
     University of Glasgow, Glasgow G12 8QQ
    \label{GLASGOW}}
\titlefoot{LPNHE, IN2P3-CNRS, Univ.~Paris VI et VII, Tour 33 (RdC),
     4 place Jussieu, FR-75252 Paris Cedex 05, France
    \label{LPNHE}}
\titlefoot{Department of Physics, University of Lund,
     S\"olvegatan 14, SE-223 63 Lund, Sweden
    \label{LUND}}
\titlefoot{Universit\'e Claude Bernard de Lyon, IPNL, IN2P3-CNRS,
     FR-69622 Villeurbanne Cedex, France
    \label{LYON}}
\titlefoot{Dipartimento di Fisica, Universit\`a di Milano and INFN-MILANO,
     Via Celoria 16, IT-20133 Milan, Italy
    \label{MILANO}}
\titlefoot{Dipartimento di Fisica, Univ. di Milano-Bicocca and
     INFN-MILANO, Piazza della Scienza 2, IT-20126 Milan, Italy
    \label{MILANO2}}
\titlefoot{IPNP of MFF, Charles Univ., Areal MFF,
     V Holesovickach 2, CZ-180 00, Praha 8, Czech Republic
    \label{NC}}
\titlefoot{NIKHEF, Postbus 41882, NL-1009 DB
     Amsterdam, The Netherlands
    \label{NIKHEF}}
\titlefoot{National Technical University, Physics Department,
     Zografou Campus, GR-15773 Athens, Greece
    \label{NTU-ATHENS}}
\titlefoot{Physics Department, University of Oslo, Blindern,
     NO-0316 Oslo, Norway
    \label{OSLO}}
\titlefoot{Dpto. Fisica, Univ. Oviedo, Avda. Calvo Sotelo
     s/n, ES-33007 Oviedo, Spain
    \label{OVIEDO}}
\titlefoot{Department of Physics, University of Oxford,
     Keble Road, Oxford OX1 3RH, UK
    \label{OXFORD}}
\titlefoot{Dipartimento di Fisica, Universit\`a di Padova and
     INFN, Via Marzolo 8, IT-35131 Padua, Italy
    \label{PADOVA}}
\titlefoot{Rutherford Appleton Laboratory, Chilton, Didcot
     OX11 OQX, UK
    \label{RAL}}
\titlefoot{Dipartimento di Fisica, Universit\`a di Roma II and
     INFN, Tor Vergata, IT-00173 Rome, Italy
    \label{ROMA2}}
\titlefoot{Dipartimento di Fisica, Universit\`a di Roma III and
     INFN, Via della Vasca Navale 84, IT-00146 Rome, Italy
    \label{ROMA3}}
\titlefoot{DAPNIA/Service de Physique des Particules,
     CEA-Saclay, FR-91191 Gif-sur-Yvette Cedex, France
    \label{SACLAY}}
\titlefoot{Instituto de Fisica de Cantabria (CSIC-UC), Avda.
     los Castros s/n, ES-39006 Santander, Spain
    \label{SANTANDER}}
\titlefoot{Inst. for High Energy Physics, Serpukov
     P.O. Box 35, Protvino, (Moscow Region), Russian Federation
    \label{SERPUKHOV}}
\titlefoot{J. Stefan Institute, Jamova 39, SI-1000 Ljubljana, Slovenia
     and Laboratory for Astroparticle Physics,\\
     \indent~~Nova Gorica Polytechnic, Kostanjeviska 16a, SI-5000 Nova Gorica, Slovenia, \\
     \indent~~and Department of Physics, University of Ljubljana,
     SI-1000 Ljubljana, Slovenia
    \label{SLOVENIJA}}
\titlefoot{Fysikum, Stockholm University,
     Box 6730, SE-113 85 Stockholm, Sweden
    \label{STOCKHOLM}}
\titlefoot{Dipartimento di Fisica Sperimentale, Universit\`a di
     Torino and INFN, Via P. Giuria 1, IT-10125 Turin, Italy
    \label{TORINO}}
\titlefoot{INFN,Sezione di Torino, and Dipartimento di Fisica Teorica,
     Universit\`a di Torino, Via P. Giuria 1,\\
     \indent~~IT-10125 Turin, Italy
    \label{TORINOTH}}
\titlefoot{Dipartimento di Fisica, Universit\`a di Trieste and
     INFN, Via A. Valerio 2, IT-34127 Trieste, Italy \\
     \indent~~and Istituto di Fisica, Universit\`a di Udine,
     IT-33100 Udine, Italy
    \label{TU}}
\titlefoot{Univ. Federal do Rio de Janeiro, C.P. 68528
     Cidade Univ., Ilha do Fund\~ao
     BR-21945-970 Rio de Janeiro, Brazil
    \label{UFRJ}}
\titlefoot{Department of Radiation Sciences, University of
     Uppsala, P.O. Box 535, SE-751 21 Uppsala, Sweden
    \label{UPPSALA}}
\titlefoot{IFIC, Valencia-CSIC, and D.F.A.M.N., U. de Valencia,
     Avda. Dr. Moliner 50, ES-46100 Burjassot (Valencia), Spain
    \label{VALENCIA}}
\titlefoot{Institut f\"ur Hochenergiephysik, \"Osterr. Akad.
     d. Wissensch., Nikolsdorfergasse 18, AT-1050 Vienna, Austria
    \label{VIENNA}}
\titlefoot{Inst. Nuclear Studies and University of Warsaw, Ul.
     Hoza 69, PL-00681 Warsaw, Poland
    \label{WARSZAWA}}
\titlefoot{Fachbereich Physik, University of Wuppertal, Postfach
     100 127, DE-42097 Wuppertal, Germany
    \label{WUPPERTAL}}
\addtolength{\textheight}{-10mm}
\addtolength{\footskip}{5mm}
\clearpage
\headsep 30.0pt
\end{titlepage}
%
\pagenumbering{arabic} 
\setcounter{footnote}{0} %
\large
\newcommand{\mm}{$\pm$}
\section {Introduction}

In the Standard Model (SM), the decay of the Higgs boson to photons 
is mediated by heavy charged particles (namely $W^\pm$ 
bosons and top-quarks). 
The corresponding branching ratio $BR(h^0 \to \gamma\gamma)$ is below 0.1\%. 
At LEP, this search is thus motivated by extensions of the SM. 
Many of the proposed models may enhance the $BR(h^0 \to \gamma\gamma)$, 
either by enlarging the $h^0\gamma\gamma$ coupling \cite{anomalous} 
or by reducing the coupling of the Higgs boson to fermions \cite{fermioph}.

We investigate the fermiophobic scenario of Two Higgs Doublets 
Models (2HDM) \cite{fermioph}, in which the lightest CP-even Higgs boson 
decays to photons and can be produced together with a $Z^0$ or a CP-odd 
Higgs boson, $A^0$.
The Higgs decay modes analysed in this paper are 
$h^0 \rightarrow \gamma\gamma$ and 
$A^0 \rightarrow b\bar{b}$ or $A^0 \rightarrow h^0Z^0$. 
However, we consider also results for $h^0 \rightarrow A^0A^0$ 
and for long-lived $A^0$, as described in \cite{our_paper}.
All LEP2 data with centre-of-mass energies above 180~GeV are analysed and 
the results reported here update those in \cite{our_paper}.

The results on $h^0Z^0$ production with $h^0 \rightarrow \gamma\gamma$ 
have been interpreted in several frameworks:
previous analyses of DELPHI data can be found in \cite{our_anom} and results 
from the other LEP experiments can be found in  \cite{others_hgg}.
In all cases, $h^0Z^0$ production with subsequent $h^0 \to \gamma\gamma$ 
decay has been used as a benchmark in the search for 
particles with SM Higgs-like couplings to bosons but no coupling to fermions.

In sections \ref{sec:theory} and \ref{sec:data}, 
the 2HDM fermiophobic scenario and the data samples used will be introduced; 
the general selection of events with isolated photons will then be presented
(in section \ref{sec:phot}), 
before going into the analysis dedicated to each of the $h^0$ boson production 
mechanisms and the different final states. The results obtained for $h^0Z^0$ 
production and $h^0A^0$ production (in sections \ref{sec:hz} and \ref{sec:ha}, 
respectively) are then combined to exclude regions in the parameter space of 
the 2HDM fermiophobic scenario, in section \ref{sec:fim}.

\section{The 2HDM fermiophobic scenario}\label{sec:theory}

General Two Higgs Doublets Models without explicit CP violation are 
characterized by five physical Higgs bosons: 
two neutral CP-even bosons ($h^0$ and $H^0$),
two charged bosons ($H^\pm$) and one neutral CP-odd boson ($A^0$).
Together with the masses, the important parameters describing 2HDMs are 
the mixing angle in the neutral CP-even sector ($\alpha$) and 
the ratio of the vacuum expectation values of the two Higgs doublets
($\tan{\beta}$).

The couplings of the Higgs doublets to fermions could be realized in different 
ways, one possibility is that only one of the doublets couples to fermions. 
The coupling of the lightest CP-even boson to a fermion pair is then 
proportional to $\cos{\alpha}$. 
If $\alpha=\pi/2$, this coupling vanishes and $h^0$ becomes a fermiophobic 
Higgs \cite{fermioph}: it decays to pairs of other Higgs bosons or 
massive gauge bosons when kinematically allowed, or to two photons in a 
large region of the parameter space.

In 2HDMs, the main mechanisms for the production of neutral Higgs bosons 
at LEP are $e^+e^- \rightarrow h^0Z^0$ and $e^+e^- \rightarrow h^0A^0$, 
both proceeding via $Z^0$ exchange.
The two processes have complementary cross-sections proportional to
$\sin^2{\delta}$ and $\cos^2{\delta}$, respectively, 
where $\delta=\alpha-\beta$. 
The $\sin{\delta}$ factor rescales the $h^0Z^0Z^0$ vertex with respect to the 
SM one.

It must be noticed that there are two different Higgs potentials
which conserve CP~\cite{fermioph} (referred to as potential A and B), 
each of them a function of seven parameters -- the masses of the Higgs bosons,
the $\alpha$ and $\beta$ angles and the sum of the squares of the 
vacuum expectation values for the two doublets. 
The choice of the potential does not affect the Higgs interactions with 
gauge bosons or fermions but leads to different Higgs-Higgs couplings and 
thus different phenomenologies. 
In particular, the decay width of $h^0 \rightarrow \gamma\gamma$, 
to which diagrams with $H^\pm$ loops significantly contribute, can be changed.

For potential A, the branching ratio of the lightest CP-even Higgs 
to two photons BR($h^0 \rightarrow \gamma\gamma$) depends on $M_{h^0}$, 
and weakly on $\sin^2{\delta}$. 
Under the assumption that $M_{H^0}~\sim~1$~TeV/$c^2$ and that $h^0$ does 
not decay to other Higgs particles, the BR($h^0 \rightarrow \gamma\gamma$)
for $\sin^2{\delta}=1$ can be obtained using the SM couplings to bosons 
and no coupling to fermions. 
For other values of $\sin^2{\delta}$, it increases with respect to this 
benchmark model.
For potential B, the BR($h^0 \rightarrow \gamma\gamma$) depends also 
on $M_{A^0}$ and $M_{H^\pm}$ and there can be large cancellations 
between the several loop contributions for some values of these parameters.
In both cases, $h^0\to A^0A^0$ is dominant when $M_{h^0}>2M_{A^0}$.

The widths for the $A^0$ tree level decays (to $f\bar{f}$, $Z^0h^0$ and 
$W^\pm H^\mp$) are independent of the potential but depend on 
$M_{A^0}$, $M_{h^0}$ and $\sin^2{\delta}$. 
For low values of $\sin^2{\delta}$, also $A^0$ becomes fermiophobic and
light $A^0$ bosons can thus be long lived.

The low values of $\sin^2{\delta}$ lead also to different mass configurations,
according to the choice of potential. 
For potential A they imply $M_{h^0} \sim 0$ and 
for potential B $M_{h^0} \sim M_{A^0}$.

\section{Data samples}\label{sec:data}

The data analysed were taken by the DELPHI detector at LEP in the years from 
1997 to 2000. The corresponding average centre-of-mass energies 
($\sqrt{s}$) and integrated luminosities  ($\cal{L}$) are shown in 
Table~\ref{tab:sl}.

\begin{table}[h]
\begin{center}
\begin{tabular}{|c||r|r|r|r|r|r|r|r|r|}
\hline
year & 1997 & 1998 & \multicolumn{4}{c|}{1999} & \multicolumn{3}{c|}{2000} \\
\hline
$\sqrt{s}$ (GeV) & 182.6 & 188.6 & 191.6 & 195.5 & 199.6 & 201.6 & 205.0 & 206.5 & 206.8 \\
$\cal{L}$ (pb$^{-1}$) & 49.3 & 153.0 & 25.1 & 76.0 & 82.7 & 40.2 & 80.0 & 59.2 & 81.8 \\
\hline
\end{tabular}
\end{center}
\caption{Average centre-of-mass energies and integrated luminosities of the 
analysed samples. The sample with $\sqrt{s}$=206.5~GeV corresponds to the data
collected after the damage on the TPC.}
\label{tab:sl}
\end{table}

In the year 2000, the centre-of-mass energies ranged from 200 GeV to 209 GeV,
while most of the luminosity was collected at around 205 GeV and 207 GeV.
In the last part of the running, DELPHI suffered irreparable
damage to one of the sectors of the main tracking device, the Time Projection 
Chamber (TPC), representing 1/12 of the acceptance.
These data were analysed separately, to isolate any systematic difference.

The DELPHI apparatus and performance are described in detail in 
\cite{delphi1,delphi2}. 
The tracking system of DELPHI consisted of the TPC 
and a Vertex Detector (VD) closest to the beam pipe and was
supplemented by extra tracking detectors, the Inner and Outer Detectors 
in the barrel region, and two Forward Chambers.
It was embedded in a magnetic field of 1.2~T, aligned parallel to the 
beam axis.
The most important subdetectors for the present analysis were the 
electromagnetic calorimeters covering different polar angle regions\footnote{
The polar angle, $\theta$,
is defined in relation to the beam axis. In all cases the complementary 
value (180$^\circ -\theta$) is also assumed.
The azimuthal angle, $\phi$, is the angle in the plane perpendicular to the 
beam direction.}: 
the luminosity monitor (STIC) for $\theta$ below 11$^\circ$, 
the Forward ElectroMagnetic Calorimeter (FEMC) between 11$^\circ$ and 
35$^\circ$ and the High density Projection Chamber (HPC) above 42$^\circ$.
The regions in between the FEMC and the HPC and the HPC intermodular
division at $\theta=90^\circ$ were equipped with hermeticity counters -- 
scintillators covered with lead, so that photons could also be tagged there.
In the azimuthal intermodular divisions of the HPC, 
at $mod(\phi,15^\circ)$=7.5$^\circ$, 
the detection of photons could be complemented by the use of 
the Hadronic CALorimeter (HCAL).
The hadronic calorimeter covered 98\% of the total solid angle, 
down to 11$^\circ$, and the whole detector was surrounded by muon drift 
chambers. 
The major hardware change with respect to the description in 
\cite{delphi2}
was the inclusion of the Very Forward Tracker \cite{vft} which extended the 
coverage of the Vertex Detector down to a polar angle of 11$^{\circ}$. 
Together with new tracking algorithms, and new alignment 
and calibration procedures, this led to an improved track
reconstruction efficiency in the forward regions of DELPHI.
The tracking algorithms for the barrel part of DELPHI were also
changed to recuperate efficiency in the damaged TPC sector.

Events corresponding to the SM processes and Higgs production signals
were fully simulated for each data set. 
The main background processes were generated with KK2f~\cite{kk} 
(for $e^+e^- \to q\bar{q}(\gamma)$), KoralZ~\cite{koralz} 
(for $e^+e^- \to \nu\bar{\nu}(\gamma)$ and $e^+e^- \to l^+l^-(\gamma)$)
and Bhwide~\cite{bhwide} (for the Bhabha scattering),
all of which include a detailed description of the initial state radiation.
These processes constitute irreducible sources of background, and thus their
accurate description is a crucial point in the analysis. In a previous paper
\cite{our_paper}, the $q\bar{q}(\gamma)$ background had been simulated with 
Pythia 6.1\cite{pythia}; 
KK2f has a more accurate description of the initial state photon radiation
and gives a 30\% increase in the cross-section for the production of 
$q\bar{q}$ with two photons at high polar angle. It also has a better 
match to the DELPHI data as shown in the next sections.

Other SM processes give smaller contributions to the studied channels.
All the \mbox{4-fermion} processes (including neutral and charged currents) 
were generated as described in \cite{wphact_d}, 
using the WPHACT generator\cite{wphact}, 
and complemented with samples of multiperipheral production of 
$e^+e^-f\bar{f}$ events generated with BDK/BDKRC \cite{gg_gen} in the
regions dominated by virtual photon collisions. The QED 
$e^+e^-\to \gamma\gamma$ process and Compton events were generated with
RADCOR \cite{BK} and TEEG \cite{compton}, respectively. The hadronization and 
fragmentation in hadronic final states was performed by Pythia \cite{pythia}.

Signal processes corresponding to the different analysed final states
were generated according to Pythia 6.1~\cite{pythia} at the different
average centre-of-mass energies, for different values of the Higgs bosons
masses (from 3 to 120 GeV/c$^2$ for the $h^0$ boson and 
10 to 170 GeV/c$^2$ fro the $A^0$ boson). 
They were cross-checked with a dedicated generator HZHA\cite{HZHA}.

All the generated data sets at the different centre-of-mass energies were 
passed through the DELPHI simulation and the same reconstruction chain as data
\cite{delphi2}.
Dedicated samples of all the above processes were used to simulate the effect 
of the damaged TPC sector. Its impact on the analyses was found to be negligible
within the statistical uncertainties.

\section{Selection of events with isolated photons}\label{sec:phot}

The selection of events with isolated photons is common to all analyses
and is described in this section.
First a general selection was applied and then isolated photons and leptons
were reconstructed. 

Only events with visible energy in the polar angle region above 20$^\circ$ 
greater than 0.1$\sqrt{s}$ were accepted. 
In addition, all events were required to contain at least one charged or 
neutral object with energy above 5 GeV in the same polar angle region.
This vetoed most of the contamination from virtual photon collision 
events.

Charged particles were classified as `good' if they had
measured momentum greater than 0.1 GeV/$c$ and their tracks were
reconstructed with impact parameters to the interaction point below 4 cm in 
the transverse plane and 4/$\sin{\theta}$~cm in the beam direction.
Energy deposits in the calorimeters unassociated to 
charged particles were required to be above 0.3~GeV to be classified as 
neutral particles.

The reconstruction of photons was done in several 
steps, starting from the showers in the electromagnetic calorimeters.
The procedure described in \cite{delphi2} was followed to identify tight 
photon candidates in the HPC. This algorithm selects
electromagnetic energy deposits with shower profiles compatible with those of 
photons. 
However, showers close to the HPC divisions were accepted as loose
photon candidates even if they failed the transverse shower profile criteria 
(and/or the longitudinal shower profile criteria if their energy was 
above 25 GeV).
In the forward region, all STIC energy deposits with polar angle
satisfying $\theta < 11^\circ$ were taken to be tight photon 
candidates\footnote{
Energy depositions below 3$^\circ$ were discarded from 
the events, to avoid contamination from off-momentum beam electrons.}. 
An algorithm was used to correct for the effects of photon conversion and 
shower development in the detector material in front of the FEMC, 
as explained below.
Electromagnetic deposits close in space were clustered together 
and the association with reconstructed charged particle tracks was used for 
electron/photon discrimination. 
Tight photon candidates were required to have no association to 
VD track elements, 
nor to signals from different combinations of other tracking 
detectors (depending on the shower polar angle).
Care was taken to exclude those tracks which were likely to come from the 
development of showers outside the calorimeter. Loose photon candidates were 
allowed to have two associated tracks.
In addition, the ratio of electromagnetic energy 
to the total energy around the cluster
-- in an angular region defined by $|\Delta{\theta}|<15^\circ$ and 
$|\Delta{\phi}| < \min(15^\circ,6^\circ \cot{\theta_{cluster}})$ --
was required to be above 90\%. 

The identification of isolated photons started from the candidates defined 
above and used a double cone centred around the photon axis, as explained 
below. Only isolated photons with energies above 5 GeV were considered. 
The selection criteria depended on the topology of the event and were as 
follows.

For the topology with photons only, 
the total energy inside a cone with half-angle of 10$^{\circ}$ was associated 
to the photon, while, to ensure isolation, the total energy between 
10$^{\circ}$ and 12$^{\circ}$ was not permitted to exceed 5~GeV. 

For the other topologies, the criteria were different for loose and tight
photon candidates. For loose photon candidates,
the half-angle of the inner cone was of 5$^{\circ}$ and the energy between 
5$^{\circ}$ and 15$^{\circ}$ was not permitted to exceed 1~GeV. 
In the case of photons tightly identified by the shower profile analysis,
no further association was done and only the external cone was kept.
The angle of the external cone, $\alpha$, was varied according to the 
energy of the photon candidate, down to 3$^{\circ}$ for $E_{\gamma}>$ 90~GeV. 
The energy limit inside the cone was rescaled to
$\sin\alpha/\sin{15^{\circ}}$~GeV, but one energetic particle was accepted 
inside this cone (and excluded from the energy calculation).

The identification of isolated photons in the barrel region required also 
that there was no HPC layer with more than 90\% of their 
electromagnetic energy, while the hadronic energy depositions above 3 GeV 
must be concentrated in  the first layer of the HCAL.

The reconstruction of isolated leptons followed the same double cone criteria,
starting from good charged particles with momentum above 4~GeV/$c$. The total 
charged particle momentum and energy deposition inside the inner cone around 
the charged particle track were associated, and an external cone used to 
ensure isolation.
Isolated charged particles associated to signals in the muon chambers, and
for which the ratio between the energy deposited in calorimeters
and the measured momentum
was less than 20\%, were identified as tight muon candidates. 
Electromagnetic showers associated to good charged particles, 
reconstructed with algorithms similar to those of photon reconstruction, 
were considered as tight or loose electron candidates.

The separation between electrons and photons converting in the tracking 
system relied on the information from the VD and its association to
isolated charged particles. 
A VD track element was defined as at least two signals in different layers
of the detector, which were associated to an isolated particle if aligned 
within 3$^\circ$ of its azimuthal direction (10$^\circ$ in the forward region).
Isolated charged particle tracks not associated to VD track elements were 
considered as candidates for photons converting in the tracking system.
At most one converted photon candidate was allowed per event, except for the 
$l^+l^-\gamma\gamma$  topologies where no recovery of converted photons was 
performed.

\section{Search for $h^0Z^0$ production}\label{sec:hz}

\subsection{Two photons and jets}\label{sec:qqgg}

In $h^0Z^0$ production, the final state with highest branching fraction is the
one corresponding to the $Z^0$ decay to hadrons, and $h^0$ decay to photons.
The preselection of this topology required that at least six good charged 
particles were present and the visible energy in the polar angle region above
20$^\circ$ was greater than 0.2$\sqrt{s}$.
All selected charged and neutral particles, except for isolated photons and 
leptons, were clustered into two jets using the DURHAM jet 
algorithm~\cite{durham}.
Two isolated photons were required in the event, and extra isolated particles 
were allowed only if their transverse momentum with respect to one of the jets 
was less than 20 GeV/$c$.

The main irreducible background to this search was $q\bar{q}$ 
production with two photons coming predominantly from initial state radiation:
at LEP2, about half of the $q\bar{q}$ events were radiative return events, 
with an effective collision energy around $M_{Z^0}$. 
Very small contributions from other sources, namely $W^+W^-$ production, 
were also present.

After the selection of the two isolated photons in the event, 
further requirements on their isolation and polar angle reduced both
Initial and Final State Radiation (ISR and FSR) contributions. 
The most energetic photon had to have at least 15\% of the 
beam energy, the minimum transverse momentum of the photons with respect 
to the jets had to be above 7.5 GeV/$c$ and the minimum polar angle 
above 30$^\circ$. 

To improve the resolution on the invariant mass of the photon pair, 
a kinematic fit imposing energy-momentum conservation and constraining 
the jet-jet invariant mass to the $Z^0$ mass was performed according to 
\cite{pufit}. 
The main inputs to the fit are the directions and energies of the photons,
which are considerably less well measured if the photons are not well 
contained in the calorimeters.
$\sigma_E$, the relative error on the measured energy $E_{\gamma i}^{meas}$, 
was obtained from calorimeter resolution studies. It was set to 100\% for 
photons reconstructed in the calorimeter boundaries, and events in which 
both photons had \mbox{$\sigma_E=100\%$} were rejected. 
For the other events, a partial $\chi^2$ was constructed as:
\begin{equation}
\frac{\chi^2_{\gamma\gamma}}{\small ndof}=\frac{1}{2}\sum_{i=1,2}
\left( \frac{E_{\gamma i}^{fit} - E_{\gamma i}^{meas}}
{\sigma_E E_{\gamma i}^{meas}} \right)^2,
\label{eq:chi}
\end{equation}
If both photons were well contained in the calorimeters, 
it was required that $\chi^2_{\gamma\gamma}/ndof$ was below 5. 
If one of the photons was not well contained, 
the information from the jets had a larger weight and
it was the global $\chi^2/ndof$ of the fit (defined in \cite{pufit})
that was required to be below 5.

\begin{figure}[H]
\begin{minipage}{.95\linewidth}
\begin{center}
{\Large DELPHI} 
\end{center}
\end{minipage}
\vfill
\begin{minipage}{.95\linewidth}
\begin{center}
\mbox{\epsfig{file=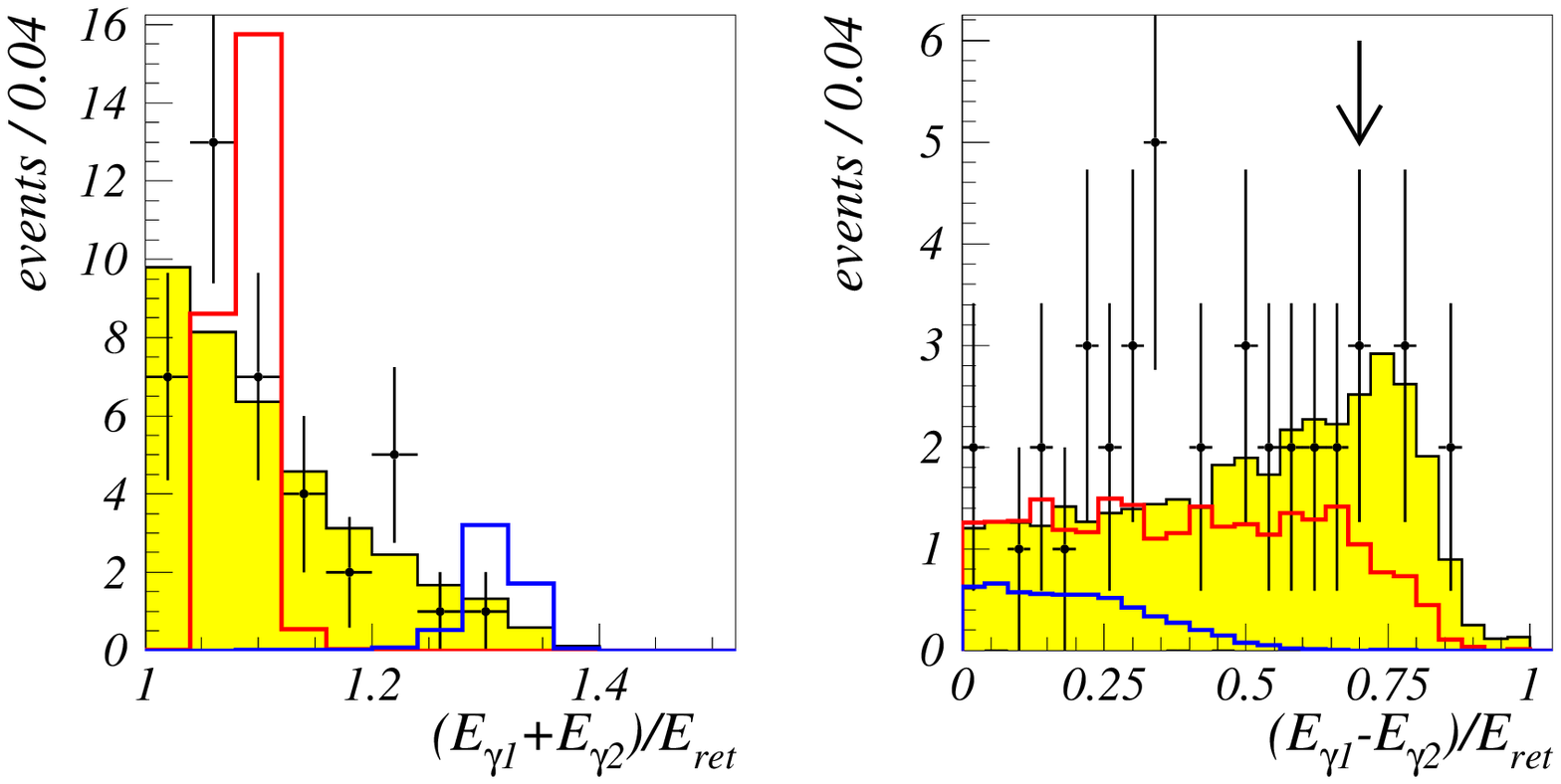,height=0.65\textheight}}
\end{center}
\vspace{-7.5cm}
\end{minipage}
\vfill
\begin{minipage}{.95\linewidth}
\mbox{\epsfig{file=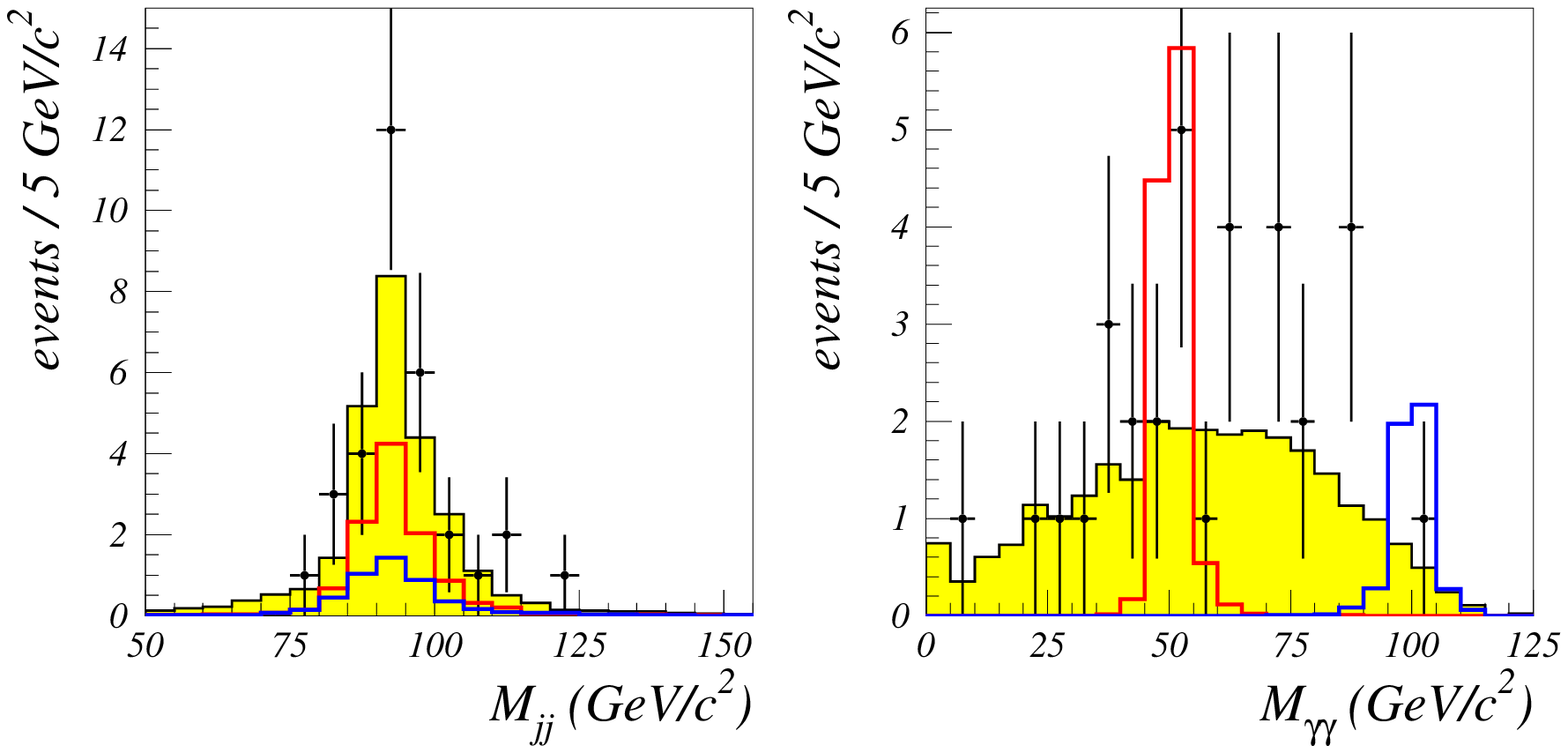,height=0.65\textheight}}
\vspace{-7.cm}
\end{minipage}
\caption{$h^0Z^0 \to \gamma\gamma q\bar{q}$. The top plots show the 
sum (left) and difference (right) of the two photon energies divided by
the radiative return energy before the cut on the energy difference 
(indicated by the arrow). The bottom plots show the reconstructed 
invariant jet-jet mass (left) and the fitted Higgs mass (right). 
The full data set (dots) is compared to the SM background (shaded area)
and two  $h^0Z^0$ signals: with $M_{h^0}$=50 GeV/$c^2$ and $M_{h^0}$=100 
GeV/$c^2$ (thick lines), shown with arbitrary normalization.}
\label{fig:qq1}
\end{figure}

At this stage, the contribution from non-$q\bar{q}(\gamma)$ events becomes 
almost negligible. 
The characteristics of the radiative return events were then used to 
reduce the background further. 
In most of these events, the photons are in the forward part of the detector 
and, usually, one of the photons carries most of the energy necessary to bring 
the $Z^0$ boson on-shell: 
\begin{equation}
E_{ret} = \frac{s-{M_{Z}}^2}{2\sqrt{s}}. 
\label{eq:eret}
\end{equation}
Only events in which the energy difference between the two photons was lower 
than 0.70$E_{ret}$ were kept.
Figure~\ref{fig:qq1} shows the sum and difference of the two photon energies
normalized to the radiative return energy.
In the final sample, one of the photons was required to be inside the HPC
acceptance.

\begin{table}[H]
\begin{center}
\begin{tabular}{|c|r|r|r|r|r|}
\hline
Selection 
& Data & Background & $q\bar{q}$ & eff$_{50}$ &eff$_{100}$\\
\hline
2 photons  & 338  & 341.5 \mm 9.5 & 296.5 \mm 9.2    &60\%&60\%\\
\hline
no-ISR/FSR & 23 &  22.9 \mm 0.3   & 21.2 \mm 0.3 &39\%&45\%\\
\hline
Z fit        & 12 &  12.4 \mm 0.2 & 12.3 \mm 0.2 &31\%&41\%\\
\hline
$\Delta E$    &  9 &   8.7 \mm 0.2 &  8.6 \mm 0.2 &26\%&40\%\\
\hline
HPC          &  9 &   8.4 \mm 0.2 & 8.3 \mm 0.2 & 26\%&40\%  \\
\hline
\end{tabular}
\end{center}
\caption[$h^0Z^0$ selection in 2000]
{$h^0Z^0 \to \gamma\gamma q\bar{q}$ selection in the year 2000: 
the evolution of the data sample with the analysis cuts is shown for the 
full 2000 data set,
compared  to the total expected background (and the $q\bar{q}$ background) 
and the efficiencies for Higgs masses of 50 GeV/$c^2$ and 100 GeV/$c^2$. 
The errors on the backgrounds are statistical only. The absolute
statistical error on the efficiencies is 1-2\%.}
\label{tab:hz_qqgg}
\end{table}

Table~\ref{tab:hz_qqgg} shows, for the data collected in the year 2000, 
the evolution with the selection criteria of the number of selected data 
events, corresponding SM background and efficiencies for two Higgs 
signals of different masses.
The contribution of the irreducible $q\bar{q}$ background is shown separately.
The numbers of selected events in all data and simulated samples
at the final selection level are shown  in Table~\ref{tab:hz_candidates}. 
The reconstructed jet-jet invariant masses and the fitted $\gamma\gamma$ 
masses are also shown in Figure~\ref{fig:qq1}.

\subsection{Two photons and two charged leptons}

The production of $h^0Z^0$ with subsequent decays $h^0 \to \gamma\gamma$ and
$Z^0 \to l^+l^-$  was analysed for all the three charged lepton flavours.
A common preselection required the presence of two isolated photons and two 
isolated leptons in events with at most five good charged 
particles and visible energy in the polar angle region above 20$^\circ$ 
greater than 0.2$\sqrt{s}$.
The most energetic photon was required to have energy greater 
than 0.1$\sqrt{s}$ and the polar angle of the leptons was required 
to be above 20$^\circ$ (except for the ones identified as muons).

The background processes for these channels include 
$e^+e^-\rightarrow Z^0/\gamma^* \rightarrow l^+l^-$ and the 
$t$-channel Bhabha scattering, which has a very large cross-section. 
To reduce the number of events with FSR, the minimum transverse momentum of 
each photon with respect to any of the leptons was required to be greater 
than 5 GeV/$c$.
The contribution from events with ISR, 
namely from radiative return to the $Z^0$, was 
reduced requiring that both photons had polar angles above 30$^\circ$, 
that there was at least one photon in the HPC, 
and imposing that the difference 
between the energies of the two photons was below 0.7$E_{ret}$ 
(see equation \ref{eq:eret}).

A kinematic fit imposing energy-momentum conservation and using the measured 
directions of the four particles in the event was performed, to rescale
the energies of leptons and photons.
Partial $\chi^2/ndof$ associated to the leptons or to the photons
were defined, as in equation \ref{eq:chi}, for the photon energies and 
lepton momenta; events were accepted only if one of the two was below 10. 
Only events for which the fitted di-lepton invariant mass was between 
60 and 120 GeV/$c^2$ were kept.

\begin{figure}[H]
\begin{minipage}{.75\linewidth}
\begin{center}
{\Large DELPHI} 
\end{center}
\end{minipage}
\vfill
\begin{minipage}{.75\linewidth}
\mbox{\epsfig{file=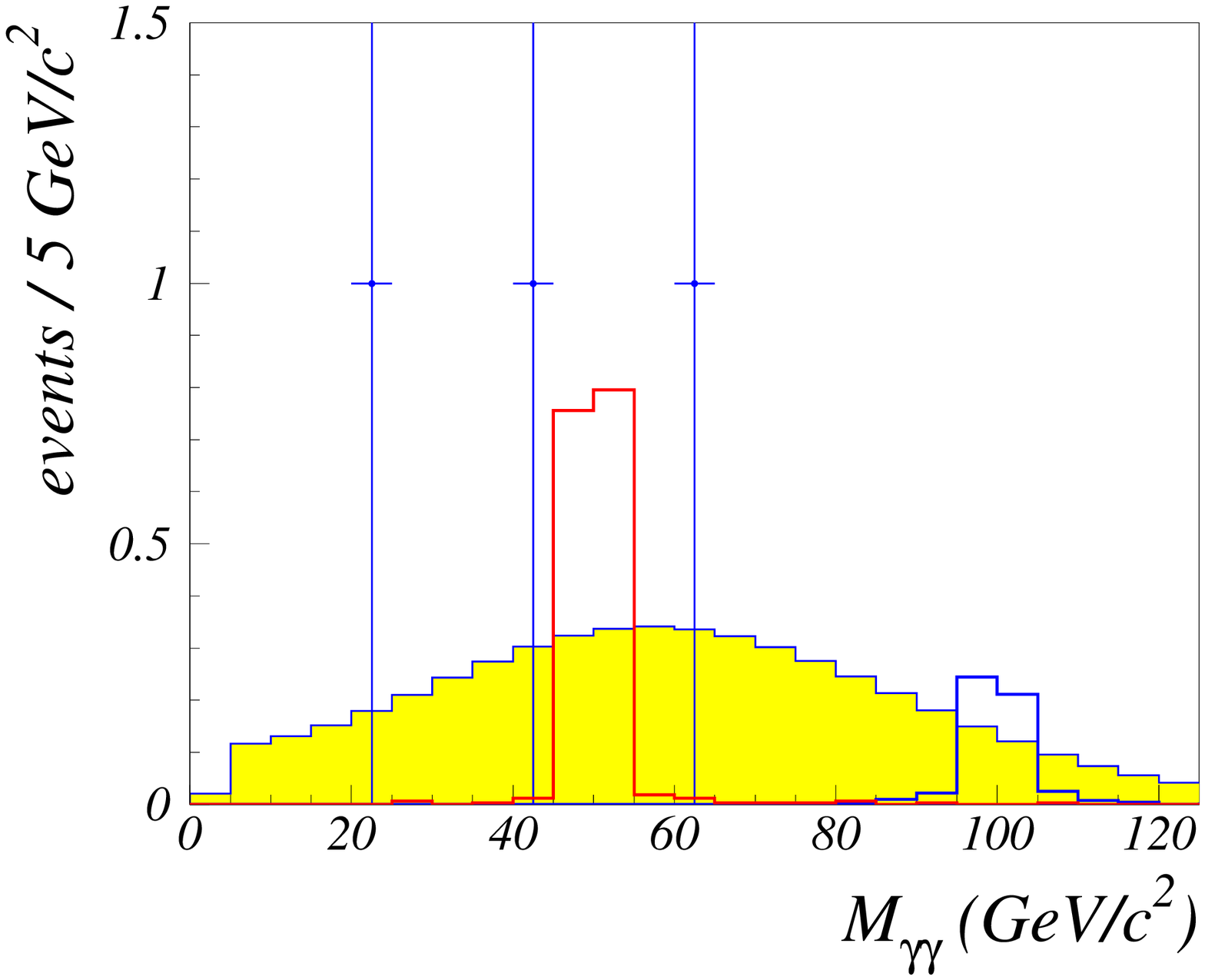,width=1.1\linewidth}}
\end{minipage}
\vfill
\begin{minipage}{.75\linewidth}
\mbox{\epsfig{file=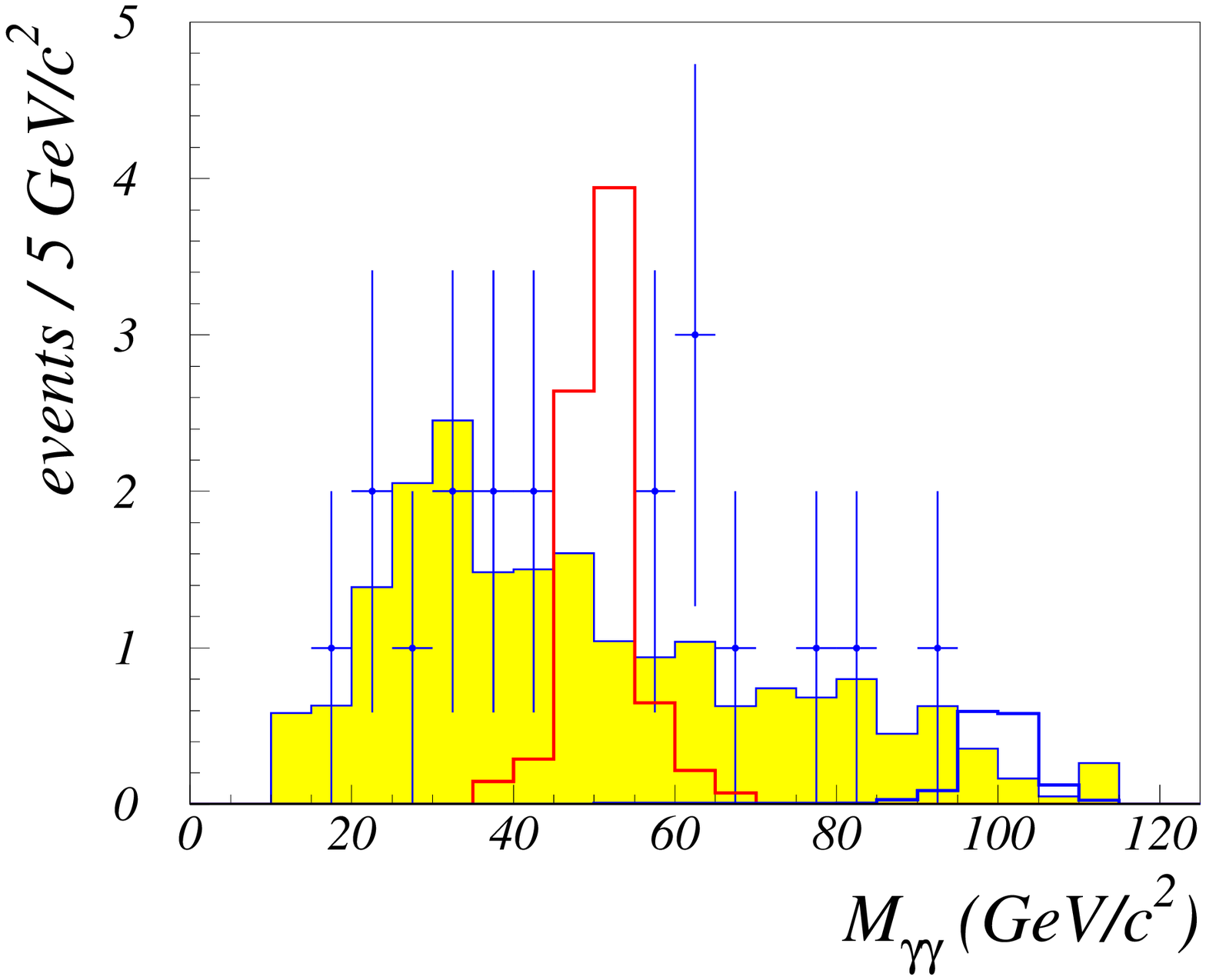,width=1.1\linewidth}}
\end{minipage}
\caption{$h^0Z^0 \to \gamma\gamma l^+l^-$ (top) and 
$h^0Z^0 \to \gamma\gamma \nu\bar{\nu}$ (bottom).
The invariant masses $M_{\gamma\gamma}$ at the last selection level for all
analysed data (dots) are compared to the SM background expectations 
(shaded histogram), and $h^0Z^0$ signals of 50 and 100 GeV/$c^2$ (top), 
shown with arbitrary normalization.}
\label{fig:ll_nn}
\end{figure}

Table~\ref{tab:ggll} presents the number of selected events in the
year 2000 data and the corresponding SM background expectations, at each
selection level, together with the efficiencies for two Higgs masses. 
The selection efficiencies for the different leptonic decays of the $Z^0$ 
vary between 13\% for $\tau^+\tau^-$ and 38\% for $\mu^+ \mu^-$ for a Higgs 
mass of 100~GeV/$c^2$. 
The final invariant mass spectra are shown in Figure~\ref{fig:ll_nn}, and 
the corresponding numbers of selected events in each data set are given in
Table~\ref{tab:hz_candidates}.

\begin{table}[H]
\begin{center}
\begin{tabular}{|c|r|r|r|r|}
\hline
Selection 
& Data & Background &eff$_{50}$ & eff$_{100}$ \\
\hline
2 photons  & 24  & 28.8 \mm 1.1 & 36\% & 36\% \\
\hline
no-ISR/FSR &  5 &   5.3 \mm 0.5 & 22\% & 31\% \\
\hline
E-p fit + $M_{Z^0}$& 0 &1.6 \mm 0.2 & 17\% & 27\% \\
\hline
\end{tabular}
\end{center}
\caption{$h^0Z^0 \to \gamma\gamma l^+l^-$ selection in the year 2000: 
the evolution of the data sample with the analysis cuts is shown together 
with the total expected background (with the corresponding statistical errors)
and the efficiencies for Higgs masses of 50 GeV/$c^2$ and 100 GeV/$c^2$
(with 1-2\% statistical error).}
\label{tab:ggll}
\end{table}

\subsection{Two photons and missing energy}

$h^0Z^0$ production with the $Z^0$ decay into neutrino pairs leads to 
purely photonic final states with large missing energy.
All purely photonic candidate events were allowed to have at most 5 good 
charged particles, none of them associated to VD track elements or 
to signals in the muon chambers. 
Charged particles not associated to energy deposits above 5~GeV had 
to have momenta below 5 GeV/c and the minimum transverse momentum of each 
individual particle with respect to the other particles reconstructed in 
the same hemisphere had to be greater than 5~GeV/c.

Cosmic rays leaving
energy deposits in the calorimeters are an important source of 
background for final states with photons and missing energy. 
Most cosmic ray events crossing the tracking system were removed by requiring 
all charged particles in the event to satisfy the impact parameter 
selection criteria defined in section \ref{sec:phot}.
Cosmic ray events crossing the detector outside the
tracking devices were vetoed by requiring that the total unassociated energy
deposition in the event was less than 10\% of the energy of the isolated 
photons. 
Events in which more than 98\% of the energy of the photon candidates was 
reconstructed from depositions in the hadronic calorimeters were 
also vetoed.
In addition, the direction of photons reconstructed in the 
HPC was required to be consistent within 25$^\circ$ with the hypothesis 
that the particle was coming from the primary vertex.

After this preselection, only events with two photons reconstructed 
with polar angles above 
25$^\circ$ and well contained either in the FEMC or in the HPC 
were kept.

The main background process for this channel is the production of 
$\nu\bar{\nu}$ pairs (either through $Z^0$ or $W^\pm$ boson exchange) 
with the emission of ISR photons. 
Even though the cross-section is low, these events constitute an irreducible 
source of background. 
The background from the QED reaction $e^+e^- \rightarrow \gamma\gamma(\gamma)$,
which has a large cross-section, was much reduced by appropriate 
kinematic requirements. 
Most of the QED events were vetoed by imposing that the angle between the two 
photons was lower than 178$^\circ$ and that the energy of the most energetic
photon was below 0.4$\sqrt{s}$.
Additional criteria were applied to events with photons in the less efficient 
regions of the electromagnetic calorimeters.
If one of the photons was either within 1.5$^\circ$ of an azimuthal 
modular division of the HPC, or its polar angle corresponded to the HPC 
edges, (defined by the intervals $[42^\circ,44^\circ]$ and 
$[87^\circ,88^\circ]$), the acoplanarity\footnote{
The acoplanarity between two objects is defined as the complement of the angle 
between them in the plane transverse to the beam direction.} 
was required to be above 3$^\circ$ and 5$^\circ$, respectively.  
If there was one converted photon in the FEMC, the acoplanarity was also 
required to be above 3$^\circ$, in order to account for the larger deflection 
of charged particles in the forward region.
Any signal observed in the hermeticity counters in between the FEMC and 
the HPC was required to be within 20$^\circ$ of a reconstructed photon.

A kinematic fit was then performed imposing the $Z^0$ mass on the invisible 
system, after requiring that the missing mass of the selected event was 
larger than 20 GeV/$c^2$. 
The global $\chi^2/ndof$ \cite{pufit}
 resulting from the fit was required to be below 5.

The results of the selection in the year 2000 data set are summarized in 
Table~\ref{tab:hz_nngg}, where the SM expectations were corrected for 
trigger efficiencies\footnote{The trigger efficiencies for events with two 
photons in the barrel part of DELPHI were of 96\%-98\%; in the forward
part, as for other final states, they were of $\sim$100\%.
In the other channels the effect of this correction is thus negligible.}. 
The invariant mass spectra at the final selection level are shown in 
Figure~\ref{fig:ll_nn}, and the corresponding numbers of selected events 
in each data set are given in Table~\ref{tab:hz_candidates}.

\begin{table}[H]
\begin{center}
\begin{tabular}{|c|r|r|r|r|r|r|}
\hline
Selection 
& Data & Background & QED & 
$\nu\bar{\nu}\gamma\gamma$ & eff$_{50}$ &eff$_{100}$\\
\hline
\multicolumn{1}{|c|}{preselection} &
1077 & 1121.2 \mm 7.9 &  1110.6 \mm 7.9 & 10.6 \mm 0.8 & 60\% & 64\% \\
\hline
\multicolumn{1}{|c|}{2 $\gamma s$ : $\theta_\gamma\; {\in}\!\!\!\slash \; [35^\circ,42^\circ]$} &
 879     & 924.5  \mm 7.2 &   915.7 \mm 7.1 & 8.8 \mm 0.7 & 50\% & 56\%\\
\hline
\multicolumn{1}{|c|}{$\alpha_{\gamma \gamma}<178^\circ$} &
 206    & 200.0  \mm 3.3 & 191.2 \mm 3.2 & 8.8 \mm 0.7 & 50\%& 56\%\\
\hline
\multicolumn{1}{|c|}{QED veto} &
 13    & 9.0  \mm 0.7 & 1.2 \mm 0.3 & 7.8 \mm 0.7 & 48\% & 53\% \\
\hline
\multicolumn{1}{|c|}{$Z fit$} &
6    & 6.6  \mm 0.6 & 0.5 \mm 0.2 & 6.1 \mm 0.6 & 44\% & 50\% \\
\hline
\end{tabular}
\end{center}
\caption{$h^0Z^0 \to \gamma\gamma\nu\bar{\nu}$ selection in the year 2000: 
the evolution of the data sample with the analysis cuts is shown, 
together with the total expected background 
(and the $QED$ and $\nu \bar{\nu} \gamma\gamma$ expectations) 
and corresponding statistical errors.
The efficiencies for Higgs masses of 50 GeV/$c^2$ and 100 GeV/$c^2$ are
shown in the last two columns, and have associated statistical errors of 
1-2\%.}
\label{tab:hz_nngg}
\end{table}

\subsection{Results on $h^0Z^0$ production}

A good agreement between data and SM expectations was found in all analysed 
channels, with a total of 54 events selected and 52\mm 1 expected. 
The numbers of selected events per channel
and centre-of-mass energy are shown in Table~\ref{tab:hz_candidates}.

\begin{table}[H]
\begin{center}
\begin{tabular}{|l|rr|rr|rr|}
\hline
$\sqrt{s}$ & 
\multicolumn{2}{c|} {$\gamma\gamma q\bar{q}$} & 
\multicolumn{2}{c|} {$\gamma\gamma l^+l^-$} &
\multicolumn{2}{c|} {$\gamma\gamma \nu\bar{\nu}$} \\
(GeV)&DATA& MC  &DATA& MC & DATA & MC \\
\hline
183 & 4 & 2.64 \mm 0.12 &-- & -- & 2 & 1.59 \mm 0.65 \\
\hline
189 & 8 & 6.85 \mm 0.33  & 1 & 1.70 \mm 0.44& 5 & 4.82 \mm 0.55\\
\hline
192 & 0 & 1.17 \mm 0.05  & 0 & 0.27 \mm 0.07& 1 & 0.93 \mm 0.13\\
\hline
196 & 4 & 3.24 \mm 0.15  & 2 & 0.62 \mm 0.15 & 2 & 2.29 \mm 0.26\\
\hline
200 & 4 & 3.15 \mm 0.15 & 0 & 0.60 \mm 0.15  & 2 & 2.31 \mm 0.24\\
\hline
202 & 3 & 1.66 \mm 0.08  & 0 & 0.25 \mm 0.07& 1 & 0.99 \mm 0.15\\
\hline
205 & 3 & 3.13 \mm 0.10  & 0 & 0.61 \mm 0.16& 2 & 2.40 \mm 0.29\\
\hline
206.5& 3 & 2.20 \mm 0.10 & 0 & 0.29 \mm 0.07 & 4 & 1.58 \mm 0.22 \\
\hline
207 & 3 & 3.11 \mm 0.14  & 0 & 0.69 \mm 0.17& 0 & 2.58 \mm 0.31\\
\hline
TOT &32 & 27.15 \mm 0.47   & 3 & 5.03 \mm 0.55 & 19& 19.49 \mm 1.06\\
\hline
\end{tabular}
\end{center}
\caption[Selected data for $h^0Z^0$ production]{Number of selected 
events for $h^0Z^0$ 
production and corresponding background expectations for the three
topologies and all data samples considered. The errors on the background
are statistical only.}
\label{tab:hz_candidates}
\end{table}

Limits on the production cross-section of $h^0Z^0$ with 
$h^0\rightarrow\gamma\gamma$, 
as a function of $M_{h^0}$, were obtained by 
combining all the channels and centre-of-mass energies using 
the Modified Frequentist Likelihood  Ratio method \cite{Read}, 
taking into account the measured and expected $M_{\gamma\gamma}$
invariant mass  distributions. 
The charged lepton flavours were added in a single channel 
for each centre-of-mass energy.

The signal samples generated with Pythia 6.1 were cross-checked with 
samples generated with HZHA and found to be compatible within
the statistical uncertainty of 1\%. 
Different fragmentation/hadronization models were also compared, and 
found to have negligible impact in the $q\bar{q}\gamma\gamma$ selection. 
Extra systematic effects could come from the parameterization of the
expected invariant $\gamma\gamma$ mass distributions. They 
were found to be small when compared to the statistical 
uncertainty on the total background expectations. 
To account for all effects, a systematic error of \mm~3\% was assigned to the 
signal efficiency used in the limit calculation.

\begin{figure}[H]
\begin{center}
\mbox{\epsfig{file=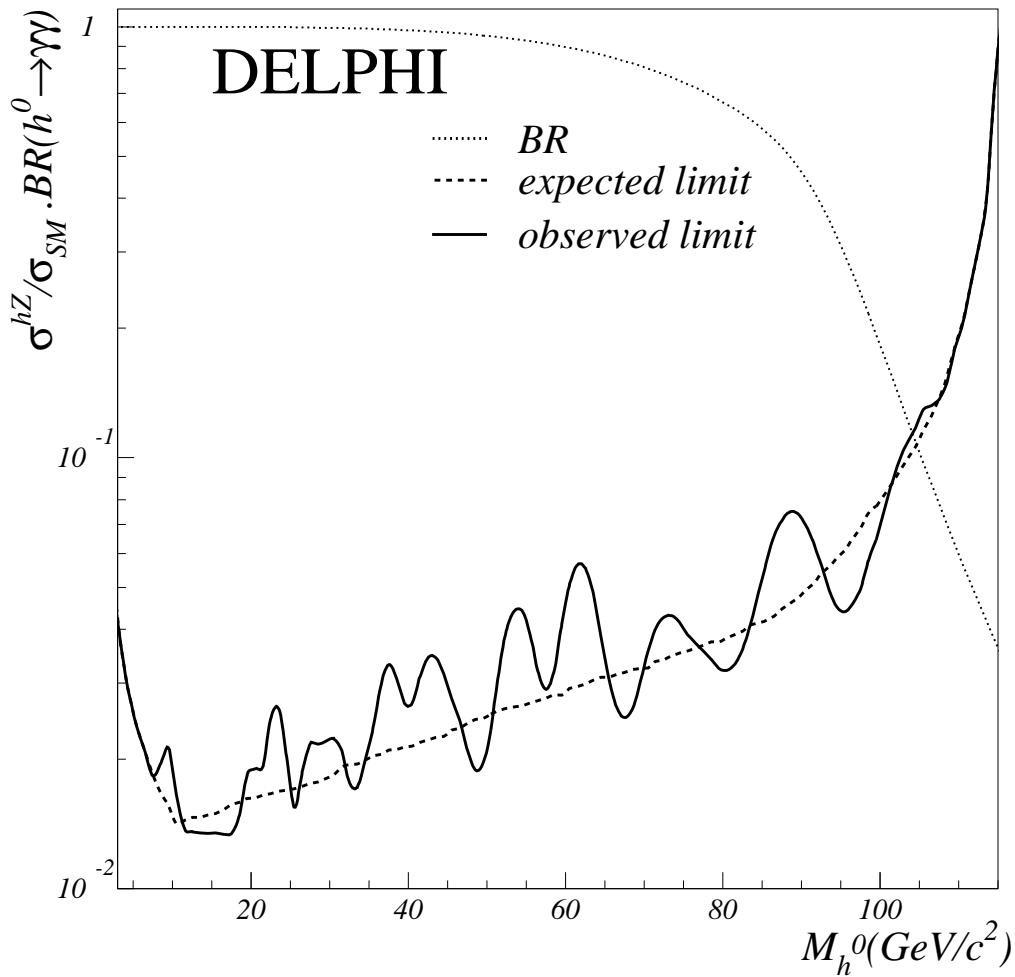,height=0.6\textheight}}
\end{center}
\vspace{-0.5cm}
\caption[95\%~CL upper limit on the $h^0Z^0$ production]{95\%~CL upper
limit on the $h^0Z^0$ production cross-section 
$\times$ BR($h^0\rightarrow \gamma\gamma$) 
normalized to the SM value.
Both the observed (full line) and the expected limits (dashed line) are shown.
Also shown is the fermiophobic $BR(h^0 \to \gamma\gamma)$ (dotted line), 
obtained by keeping the SM couplings of the Higgs to boson pairs and
setting the $h^0f\bar{f}$ couplings to 0. 
A 95\%~CL mass limit is given by the intersection of 
the cross-section limit and the $BR(h^0 \to \gamma\gamma)$ 
curve at 104.1~GeV/$c^2$ 
(104.6 GeV/$c^2$ expected).}
\label{fig:hz_lim}
\end{figure}

Figure~\ref{fig:hz_lim} shows, as a function of the Higgs boson mass,
the 95\% Confidence Level (CL) upper limit on $BR(h^0 \to \gamma\gamma)$
times the ratio of the $h^0Z^0$ production cross-section to the
SM one. This ratio is equal to $\sin^2{\delta}$ in 2HDMs 
(see section \ref{sec:theory}). 
For the range of masses studied, and taking the model in \cite{fermioph} as 
a reference, the width of the Higgs boson is always below the mass resolution 
of the analyses\footnote{For $M_{h^0}=100$~GeV/$c^2$, the mass resolutions 
were of 1\% for the leptonic channel, 2.5\% in the hadronic channel, and 
3.5\% in the missing energy channel.}.

In a model where the Higgs couplings to bosons have SM values, but the 
couplings  to fermions vanish, a 95\%~CL lower limit on the $h^0$ boson mass
is given by the intersection of the cross-section limit and the prediction for
$BR(h^0 \to \gamma\gamma)$ (also shown in Figure~\ref{fig:hz_lim}), 
at 104.1 GeV/$c^2$. 
The expected limit in this case is 104.6 GeV/$c^2$.

\section{Search for $h^0A^0$ production}\label{sec:ha}

\subsection{Two photons and two b-jets}

The search for $h^0A^0 \to \gamma\gamma b\bar{b}$ is very similar to
the one for $h^0Z^0 \rightarrow \gamma\gamma q\bar{q}$ 
(see section \ref{sec:qqgg}). 
Similar criteria were applied, differing mainly in the kinematic fit 
performed.
In the $h^0A^0$ case less constraints were used: the jet-jet invariant mass 
was left free and an additional ISR photon was allowed in the beam direction. 
In fact, if $M_{h^0}+M_{A^0} < M_{Z^0}$, a significant part of the signal 
events may correspond to radiative returns to the $Z^0$ with subsequent 
$Z^0 \rightarrow h^0A^0$.
Since $M_{A^0}$ has to be fitted without constraints, 
the global $\chi^2/ndof$ becomes an important selection criterion:
it was required to be below 10 in all cases. 
Tighter requirements on partial $\chi^2_{\gamma\gamma}/ndof$ and
global $\chi^2/ndof$ were still apllied as discussed for the $h^0Z^0$ 
case.

\begin{figure}[H]
\vspace{-1.cm}
\begin{minipage}{1.\linewidth}
\begin{center}
{\Large DELPHI} 
\end{center}
\end{minipage}
\begin{minipage}{1.\linewidth}
\mbox{\epsfig{file=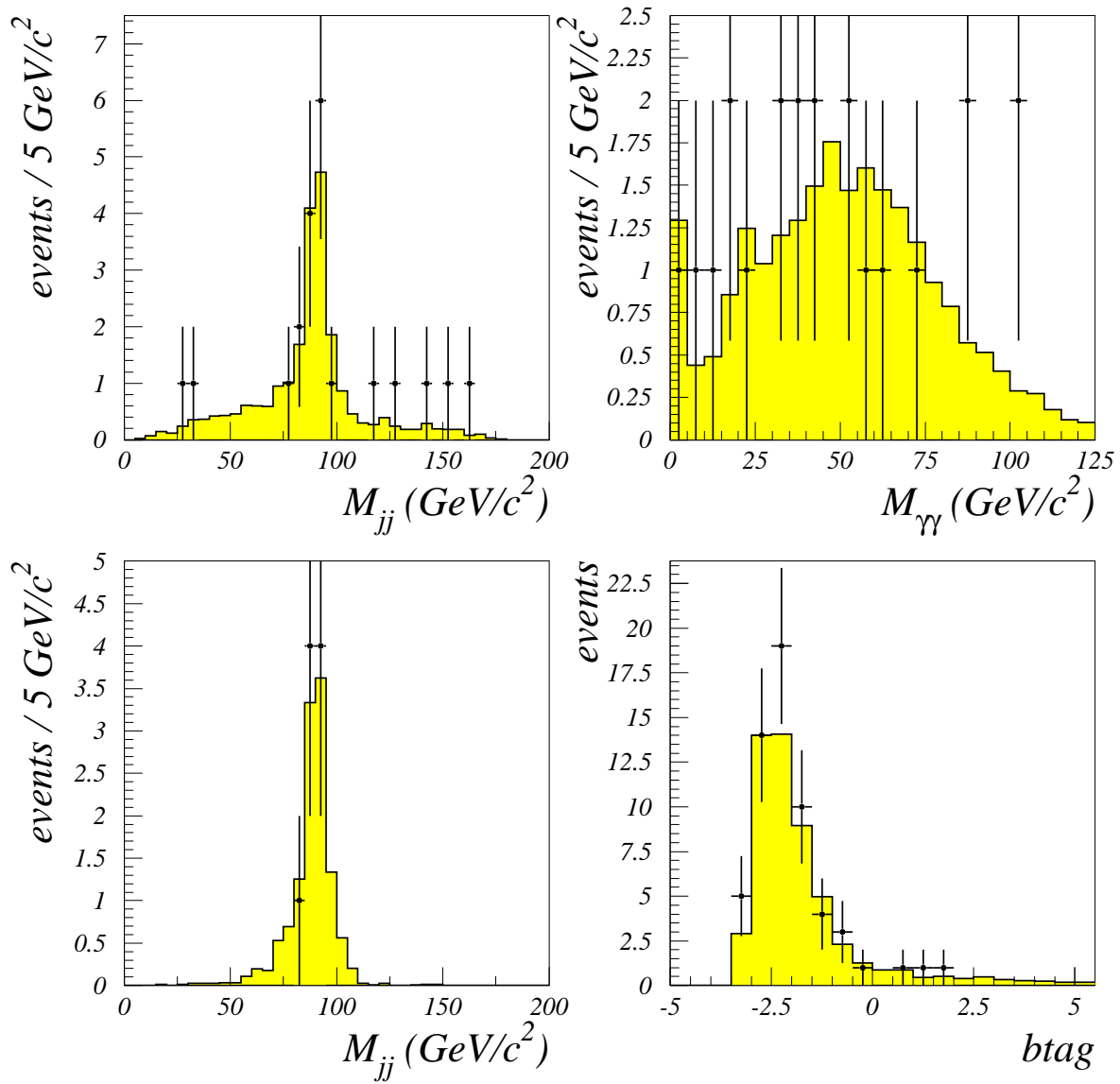,height=0.7\textheight}}
\end{minipage}
\caption{$h^0A^0 \rightarrow \gamma\gamma b\bar{b}$ analysis. 
The reconstructed $A^0$ and $h^0$ masses at the final selection level
are shown in the top plots for data and expected SM background in all
data sets.
The lower plots show the fitted jet-jet mass for events selected in both 
the $h^0A^0$ and $h^0Z^0$ analyses and the distribution of the b-tag variable
just before the cut at -2.}
\label{fig:bbgg}
\end{figure}

In the next stage, it was required that the energies of the two photons 
satisfied 
\mbox{$|E_{\gamma 1}-E_{\gamma 2}| < 0.70|E_{\gamma 1}+E_{\gamma 2}|$}. 
Again, at least one of the photons had to be in the HPC and
 the combined b-tagging of the event, as
defined in \cite{btag}, was required to be above -2.
Figure~\ref{fig:bbgg} shows the agreement of the b-tagging variable in data and
SM simulation, together with the mass distributions obtained for the $b\bar{b}$
and $\gamma\gamma$ pairs.

\begin{table}[H]
\begin{center}
\begin{tabular}{|c|r|r|r|r|}
\hline
Selection & Data & Background & $q\bar{q}$ & eff$_{100}$      \\ 
\hline
no-ISR/FSR
           & 23 & 22.9\mm 0.3 & 21.2\mm 0.3 & 53\%     \\
\hline
fit       
           & 19 & 20.1\mm 0.3 & 19.1\mm 0.3 & 49\%     \\
\hline
$\Delta E$
	  & 14 & 17.1\mm 0.3 & 16.1\mm 0.2 & 49\%     \\
\hline
HPC       
	  & 14 & 16.5\mm 0.3 & 15.5\mm 0.3 & 49\%     \\
\hline
b-tag
	  &  7 &  6.9\mm 0.2 &  6.4\mm 0.2 & 46\%     \\
\hline
\end{tabular}
\end{center}
\caption{$h^0A^0 \to \gamma\gamma b\bar{b}$ analysis: the
comparison of data and MC events 
selected at each analysis level for the year 2000 data. 
The $q\bar{q}(\gamma\gamma)$ contribution is shown separately. The
efficiency for the selection of a $h^0A^0$ signal with $M_{h^0}=100$~GeV/$c^2$
and $M_{A^0}=M_{Z^0}$ produced at $\sqrt{s}$=206~GeV is shown in the last 
column (the corresponding statistical error being of 1-2\%).}
\label{tab:ZandA}
\end{table}

The total numbers of selected data and background events 
in the analysis of the year 2000 data are shown in Table~\ref{tab:ZandA}. 
The non-$q\bar{q}$ background is larger compared to the case with the 
$Z^0$ mass constraint: it consists mainly of semi-leptonic $W^+W^-$ events.
Table~\ref{tab:ZandA} shows also the efficiency obtained for 
$e^+e^- \rightarrow h^0A^0$ 
with $M_{A^0}=M_{Z^0}$ and $M_{h^0}=100$ GeV/$c^2$.
The numbers of selected events at the final analysis level are given in
Table~\ref{tab:ha_candidates}.

\subsection{Three or more photons and jets}

Within the framework of general 2HDM, the main decay channel of $A^0$ is 
$b\bar{b}$. However, when kinematically allowed, the decay to $h^0Z^0$ becomes 
dominant in a large region of parameter space. 
This can lead to final states with a fermion pair and four photons. 
Only the $Z^0$ decay to quark pairs, which has the highest branching ratio, 
was analysed.
At least three isolated photons were required and at most one converted photon 
was allowed.

Two of the photons had to be above 30$^\circ$ in polar angle and 
have a transverse momentum with respect to any of the jets greater than 
7.5 GeV/c.  
The jet-jet invariant mass was required to be in the range 50 to 130 GeV/$c^2$.
The difference between the energy of any of the photons and the energy carried 
by all the others had to be below 0.33$E_{ret}$ 
(defined in equation \ref{eq:eret}).
Finally, one of the photons was required to be in the
HPC, and at least three photons were required to be above 15$^\circ$ in polar 
angle.

The numbers of selected events in the 2000 data sample and the corresponding 
SM expectations are given in Table~\ref{tab:qqgggg}.
The efficiencies for the signal were as high as 60\% for high masses, and  
above 40\% for most of the considered range. 
Since only three photons were required, no mass reconstruction was attempted 
and the result is based on the numbers of selected events at this stage 
(given in Table~\ref{tab:ha_candidates}).

\begin{table}[H]
\begin{center}
\begin{tabular}{|c|r|r|r|r|}
\hline
Selection & Data & Background & $q\bar{q}$ & eff \\
\hline
no-ISR/FSR &  6 &  4.22 \mm 0.14 & 4.00 \mm 0.14 & 66\% \\
\hline
$Z^0$ mass &  3 &  3.31 \mm 0.13 & 3.07 \mm 0.12 & 66\% \\
\hline
$\Delta E$ &  2 &  1.85 \mm 0.10 & 1.65 \mm 0.08 & 64\% \\
\hline
HPC        &  1 &  0.86 \mm 0.07 & 0.73 \mm 0.06 & 62\% \\
\hline
\end{tabular}
\end{center}
\caption[$h^0A^0$ analysis]{
$h^0A^0 \rightarrow q\bar{q}\gamma\gamma\gamma(\gamma)$ analysis: the
comparison of selected data and MC at each analysis level for the year 
2000 data is shown. The
efficiency for the selection of a $h^0A^0$ signal with $M_{h^0}=35$~GeV/$c^2$
and $M_{A^0}=135$~GeV/$c^2$ produced at $\sqrt{s}$=206~GeV is shown in the 
last column (the corresponding statistical error being of 1-2\%).}
\label{tab:qqgggg}
\end{table}

\subsection{Results on $h^0A^0$ production}

The numbers of selected events in each data set and for both 
channels are shown in Table~\ref{tab:ha_candidates}.
Limits were extracted with the same algorithm as for $h^0Z^0$.
In the $b\bar{b}\gamma\gamma$ channel the reconstructed mass values were
used, while for the $q\bar{q}\gamma\gamma\gamma(\gamma)$ there was no attempt 
to reconstruct the masses, and a pure counting experiment was performed.
Searches in both channels had similar sensitivity to the signal.

To extract upper 
limits on $\cos^2{\delta}$ (defined in section \ref{sec:theory}), 
reference cross-sections computed with HZHA\cite{HZHA} were used.
For the $h^0 \to \gamma\gamma$ decay, the branching ratio computed with 
$\sin^2{\delta}=1$ 
and shown in Figure~\ref{fig:hz_lim} was used. This represents a conservative 
assumption since BR($h^0 \rightarrow \gamma\gamma$) increases for lower
$\sin^2{\delta}$.
Depending on the point in parameter space under consideration, the limit was 
derived assuming a 100\% branching fraction for the dominant decay channel of 
the $A^0$ ($b\bar{b}$ or $h^0Z^0$). The limits are almost independent of the 
other model parameters, if the masses $M_{H^0}$ and $M_{H^\pm}$ are high enough
not to open new decay channels as  $A^0 \to H^\pm W^\mp$.

\begin{table}[H]
\begin{center}
\begin{tabular}{|l|rr|rr|}
\hline
$\sqrt{s}$ & 
\multicolumn{2}{c|} {$b\bar{b}\gamma\gamma$} & 
\multicolumn{2}{c|} {$q\bar{q}\gamma\gamma\gamma(\gamma)$} \\
(GeV)&DATA& MC  &DATA& MC \\
\hline
183 & 2 & 1.90 \mm 0.10 & 0 & 0.23 \mm 0.04 \\
\hline
189 & 3 & 5.82 \mm 0.30  & 0 & 0.74 \mm 0.11 \\
\hline
192 & 1 & 0.94 \mm 0.05  & 0 & 0.13 \mm 0.02 \\
\hline
196 & 4 & 2.82 \mm 0.14  & 1 & 0.29 \mm 0.05 \\
\hline
200 & 3 & 2.91 \mm 0.15 & 1 & 0.35 \mm 0.05  \\
\hline
202 & 1 & 1.50 \mm 0.07  & 0 & 0.16 \mm 0.02 \\
\hline
205 & 6 & 2.43 \mm 0.09  & 1 & 0.32 \mm 0.04 \\
\hline
206.5& 1 & 1.85 \mm 0.09 & 0 & 0.20 \mm 0.03 \\
\hline
207 & 0 & 2.59 \mm 0.10  & 0 & 0.34 \mm 0.05 \\
\hline
TOT  &21 & 22.76 \mm 0.42   & 3 & 2.75 \mm 0.15 \\
\hline
\end{tabular}
\end{center}
\caption[Selected data for $h^0A^0$ production]{Selected events for $h^0A^0$ 
production and corresponding background expectations for the two
topologies and all data samples considered. The errors on the background
are statistical only.}
\label{tab:ha_candidates}
\end{table}

\begin{figure}[H]
\begin{center}
\mbox{\epsfig{file=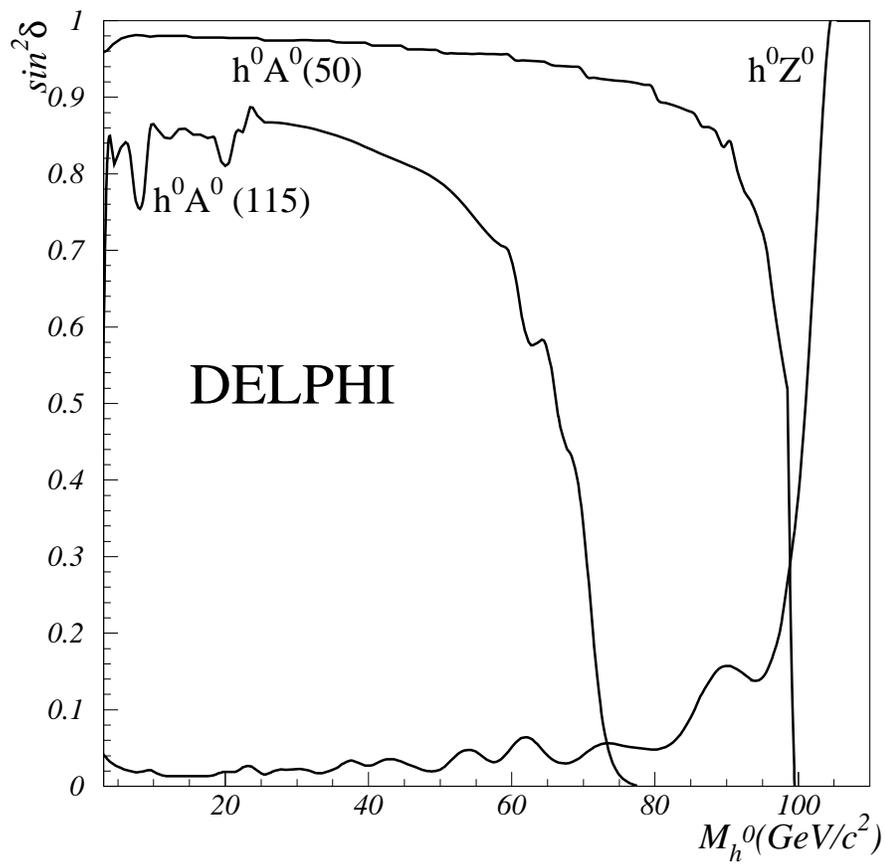,width=.8\linewidth}}
\end{center}
\caption{95\%~CL limits on the $h^0Z^0$ and $h^0A^0$ production cross-sections 
expressed in terms of $\sin^2{\delta}$.
Values above the $h^0Z^0$ curve and below the $h^0A^0$ curves 
(shown for two different $M_{A^0}$ values: 50~GeV/$c^2$ and 115~GeV/$c^2$) 
are excluded.}
\label{fig:hz+ha}
\end{figure}

Figure~\ref{fig:hz+ha} shows the limits on $\sin^2{\delta}$ as a function 
of $M_{h^0}$, obtained for two different $M_{A^0}$ values -- they strongly
depend on the mass hypothesis used for $A^0$.

\section{Exclusion in the 2HDM parameter space}\label{sec:fim}

The combination of the results on $h^0Z^0$ and $h^0A^0$ production
is illustrated in Figure~\ref{fig:hz+ha}. 
The upper limits on $\sin^2{\delta}$ for a given $M_{h^0}$ and the upper 
limits on $\cos^2{\delta}$ for a given ($M_{h^0}$,$M_{A^0}$) pair are 
combined to exclude regions in the ($M_{h^0}$,$M_{A^0}$) plane for 
all $\delta$ values.

\begin{figure}[H]
\begin{center}
\mbox{\epsfig{file=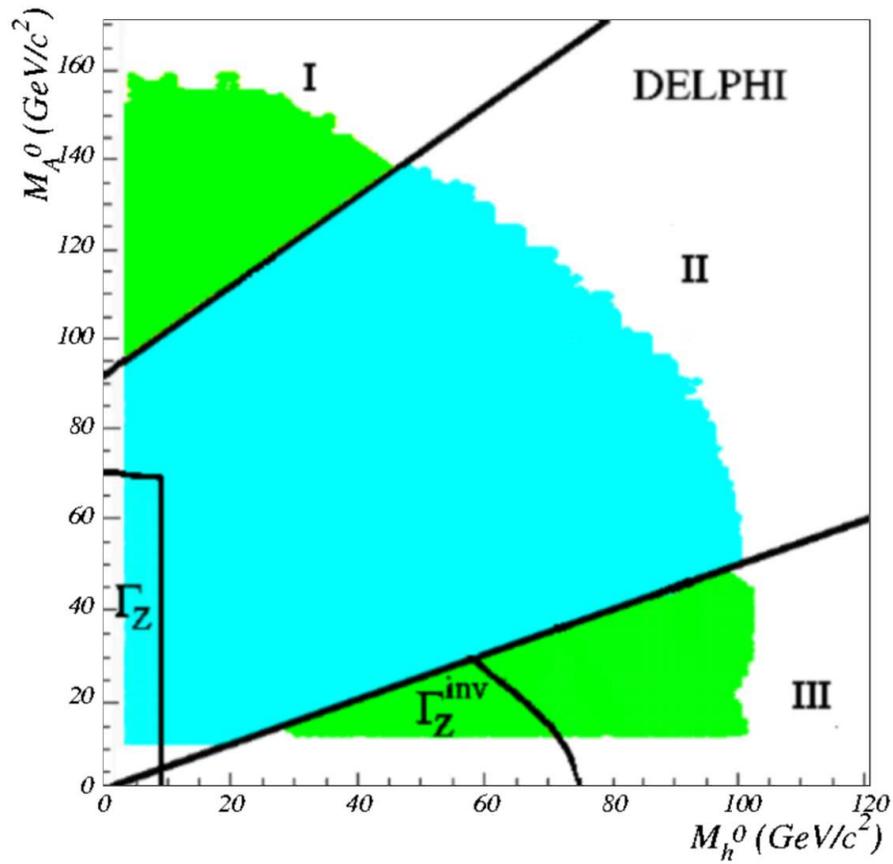,width=.8\linewidth}}
\end{center}
\caption{The shaded areas correspond to regions excluded
at 95\%~CL for all $\delta$ values.
The plot is divided into regions according to the dominant decay 
modes of $h^0$ and $A^0$, as explained in the text. The exclusions from LEP 1 
data, based on the total and invisible width of the $Z^0$, are also shown.}
\label{fig:mhma}
\end{figure}

Figure~\ref{fig:mhma} shows the 95\%~CL exclusion in the plane
$(M_{h^0},M_{A^0})$ divided into separate regions according to the 
different kinematics:

\vspace{.25cm}
{\bf I. $M_{A^0} > M_{h^0} + M_{Z^0}$}

The main decay channels in this region are
$h^0 \to \gamma\gamma$ and $A^0 \to h^0Z^0$. Thus
the $\gamma\gamma Z^0$ and $q\bar{q}\gamma\gamma\gamma(\gamma)$ topologies are 
the relevant ones.
The vertex $A^0h^0Z^0$ is proportional to $\cos{\delta}$ and thus 
the decay $A^0 \to h^0Z^0$ is dominant only for $\sin^2{\delta}<0.95$.
However, for higher $\sin^2{\delta}$ values, these mass combinations 
are excluded by the limits on $h^0Z^0$ production. 

\vspace{.25cm}
{\bf II. $M_{h^0}/2 < M_{A^0} < M_{h^0} + M_{Z^0}$}

$h^0 \to \gamma\gamma$ and $A^0 \to b\bar{b}$ are the dominant decays in
most of the parameter space. However,
the decay $A^0 \rightarrow b\bar{b}$ is suppressed as 
$\sin^2{\delta}$ approaches 0, and the $A^0$ becomes stable, and thus
invisible, for $\sin^2{\delta} < 10^{-6}$. 
The exclusion in this region comes from the search for $h^0A^0$ 
production, the results of a $\gamma\gamma + E_{mis}$ analysis
(described in \cite{our_paper}) being used.
This allows all the mass region excluded by the combination of the 
$h^0Z^0$ and $b\bar{b}\gamma\gamma$ analyses for higher $\sin^2{\delta}$ 
values also to be excluded for low $\sin^2{\delta}$. 

\vspace{.25cm}
{\bf III. $2 M_{b} < M_{A^0} < M_{h^0}/2$}

The decay $h^0 \rightarrow A^0A^0$, kinematically allowed, becomes dominant. 
The $A^0$ still decays to $b\bar{b}$, as long as $2 M_{b} < M_{A^0}$, and 
this gives rise to 6-fermion final states.
This decay is relevant also for non-fermiophobic 2HDMs, and the results 
published by DELPHI in \cite{maarten} are used. 
For very low $\sin^2{\delta}$ there is a region of totally 
invisible final states for which the results on the invisible $Z^0$ width,
obtained at LEP1, are used, as explained in \cite{our_paper}.

\vspace{.50cm}

The LEP1 results on the total $Z^0$ width
are also used to cover the region where the two masses are below the reach 
of the present analysis \cite{our_paper}.

These exclusions are valid for all the allowed parameter space in the case 
of potential A, but only for $M_{H^+}>$500~GeV/$c^2$ or $\sin^2{\delta}>0.1$
in the case of potential B. For lower values of both $M_{H^+}$ and 
$\sin^2{\delta}$, the BR($h^0 \to \gamma\gamma$) can vanish due to 
cancellation of the different loop contributions.

For the two potentials considered, the exclusions obtained for invisible $A^0$ 
(which are more restrictive than the ones for visible final states, 
in region III) apply only to small mass regions. This is because
$\sin^2{\delta}<10^{-6}$ implies $M_{h^0} \sim 0$ in potential A and a band of
$M_{h^0} \sim M_{A^0}$ in potential B. Both mass bands are in principle 
outside region III but, since their precise width is not known, the region of 
exclusion for invisible $A^0$ is also shown.

\section{Conclusions}

Around 650 pb$^{-1}$ of LEP2 data collected by DELPHI, 
at centre-of-mass energies
between 183 and 209 GeV, were analysed in the search for Higgs bosons 
decaying into photons.
In the context of 2HDM, both $h^0Z^0$ and $h^0A^0$ production were searched 
for, and a large variety of final states involving photons and fermions was
considered.
No evidence for new physics was found.

Lower limits were set on the mass of a particle with Higgs-like couplings to 
bosons and decaying to two photons. 
In a model where the Higgs couplings to bosons have SM values, but the 
couplings  to fermions vanish, a 95\%~CL lower limit on the $h^0$ boson mass
is set at 104.1 GeV/$c^2$; the expected limit in this case is 104.6 GeV/$c^2$.
Exclusions at 95\%~CL were also derived in the  mass plane $(M_{h^0},M_{A^0})$
of the fermiophobic 2HDM scenario.

\subsection*{Acknowledgements}
\vskip 3 mm
 We are greatly indebted to our technical 
collaborators, to the members of the CERN-SL Division for the excellent 
performance of the LEP collider, and to the funding agencies for their
support in building and operating the DELPHI detector.\\
We acknowledge in particular the support of \\
Austrian Federal Ministry of Education, Science and Culture,
GZ 616.364/2-III/2a/98, \\
FNRS--FWO, Flanders Institute to encourage scientific and technological 
research in the industry (IWT), Belgium,  \\
FINEP, CNPq, CAPES, FUJB and FAPERJ, Brazil, \\
Czech Ministry of Industry and Trade, GA CR 202/99/1362,\\
Commission of the European Communities (DG XII), \\
Direction des Sciences de la Mati$\grave{\mbox{\rm e}}$re, CEA, France, \\
Bundesministerium f$\ddot{\mbox{\rm u}}$r Bildung, Wissenschaft, Forschung 
und Technologie, Germany,\\
General Secretariat for Research and Technology, Greece, \\
National Science Foundation (NWO) and Foundation for Research on Matter (FOM),
The Netherlands, \\
Norwegian Research Council,  \\
State Committee for Scientific Research, Poland, SPUB-M/CERN/PO3/DZ296/2000,
SPUB-M/CERN/PO3/DZ297/2000, 2P03B 104 19 and 2P03B 69 23(2002-2004)\\
FCT - Funda\c{c}\~ao para a Ci\^encia e Tecnologia, Portugal, \\
Vedecka grantova agentura MS SR, Slovakia, Nr. 95/5195/134, \\
Ministry of Science and Technology of the Republic of Slovenia, \\
CICYT, Spain, AEN99-0950 and AEN99-0761,  \\
The Swedish Natural Science Research Council,      \\
Particle Physics and Astronomy Research Council, UK, \\
Department of Energy, USA, DE-FG02-01ER41155. \\
EEC RTN contract HPRN-CT-00292-2002. \\



\newpage

\newpage

%


\newpage

%


\end{document}